\documentclass[useAMS]{mn2e}
\usepackage{graphicx}
\newcommand\gro{GRO~J1008-57}
\title[Suzaku Observation of GRO~J1008-57]{Suzaku observation of the
transient X-ray pulsar GRO~J1008-57}

\author[Naik et al.]{Sachindra Naik$^{1}$\thanks{snaik@prl.res.in}, Biswajit Paul$^{2}$\thanks{bpaul@rri.res.in}, Chetan Kachhara$^{3}$, Santosh V. Vadawale$^{1}$\thanks{santoshv@prl.res.in}\\
$^1$Astronomy and Astrophysics Division, Physical Research Laboratory, Navrangapura, Ahmedabad - 380009, Gujarat, India\\
$^2$Raman Research Institute, Sadashivnagar, C. V. Raman Avenue, Bangalore 560080, India\\
$^3$Department of Physics, Kohima Science College, Jotsoma, Kohima - 797002, Nagaland, India}

\begin{document}

\date{}

\maketitle

\begin{abstract}
We report the timing and broad-band spectral properties of the Be transient 
high mass X-ray binary pulsar \gro~ using a $Suzaku$ observation in the 
declining phase of its 2007 November-December outburst. Pulsations with 
a period of 93.737 s were clearly detected in the light curves of the pulsar 
up to the 80-100 keV energy band. The pulse profile was found to be strongly 
energy dependent, a double peaked profile at soft X-ray energy bands ($<$8 keV) 
and a single peaked smooth profile at hard X-rays. The broad-band energy 
spectrum of the pulsar, reported for the first instance in this paper, is 
well described with three different continuum models viz. (i) a high energy 
cut-off power-law, (ii) a Negative and Positive power-law with EXponential 
cut-off (NPEX), and (iii) a partial covering power-law with high energy
cut-off. Inspite of large value of absorption column density 
in the direction of the pulsar, a blackbody component of temperature 
$\sim$0.17 keV for the soft excess was required for the first two continuum 
models. A narrow iron K$_\alpha$ emission line was detected in the pulsar 
spectrum. The partial covering model, however, is found to explain the phase 
averaged and phase resolved spectra well. The dip like feature in the pulse 
profile can be explained by the presence of an additional absorption component 
with high column density and covering fraction at the same pulse phase. 
The details of the results are described in the paper.

\end{abstract}


\section{Introduction}

High mass X-ray binary (HMXB) systems are strong X-ray emitters via the
accretion of matter from the OB companion. The majority of the HMXBs are 
known to be Be X-ray binaries. The mass donor in the Be binary systems 
is generally a B star that is still on the main sequence and lying well 
inside the Roche surface. In these Be binary systems, the compact object 
(a neutron star) is typically in a wide orbit with moderate eccentricity 
with orbital period in the range of 16 - 400 days (Coe 2000). The neutron 
star in these Be systems spends most of the time far away from the 
circumstellar disk surrounding the Be companion. Mass transfer from the 
Be companion to the neutron star takes place through the circumstellar 
disk. Strong X-ray outbursts are normally seen when the neutron star 
(pulsar) passes through the circumstellar disk or during the periastron 
passage (Okazaki \& Negueruela 2001). The outbursts in the Be/X-ray
binaries are also theoritically explained by invoking the truncation of 
the circumstellar disk (Okazaki et al. 2002). According to the model, the 
neutron star exerts a negative torque outside some critical radius  
resulting in the truncation of the Be disk. The disk matter then accumulates 
in the outer rings of the disk until the truncation is overcome by the effects 
of global one-armed oscillations, disk warping, etc. The subsequent sudden 
infall of the high-density disk matter onto the neutron star causes X-ray 
outbusts in these systems.

With the exception of a very few peculiar cases like LS~I+61303 (Massi et 
al. 2004 and references therein), all of the Be/X-ray binary systems appear 
to be accretion powered X-ray pulsars. The pulse period of the X-ray pulsars 
in Be binary systems is in the wide range of seconds to hundreds of seconds.  
There is a strong correlation between the pulse period and the orbital 
period (Corbet 1986) suggesting effective transfer of angular momentum 
from the accreted material. On the longer time-scale (months to years), 
the variability is observed in optical and infrared bands that is 
attributed to the structural changes in the circumstellar disk (Reig et
al. 2001 and references therein). The X-ray spectra of Be/X-ray binary
pulsars are usually hard. A fluorescent iron emission line at 6.4 keV is 
observed in the spectrum of most of the X-ray pulsars. It is possible that 
most of these systems have a soft X-ray excess above the power-law continuum
component. However, detection of the the soft excess depends on the value of
absorption column density (Paul et al. 2002; Naik \& Paul 2004a, 2004b and 
references therein).

The transient X-ray pulsar \gro~ was discovered on 1993 July 14 by the 
BATSE experiment onboard the {\it Compton Gamma Ray Observatory (CGRO)}
(Stollberg et al. 1993). X-ray pulsations of 93.587 s were detected in 
the 20-120 keV energy range of BATSE. From ASCA observation,
the X-ray pulse profile of the pulsar was found to have a double-peak 
structure with a well-defined, sharp intensity minimum and a less 
prominent secondary minimum (Tanaka et al. 1993).The BATSE spectrum was
described by an optically thin thermal bremsstrahlung model with {\it kT} 
= 25 keV. Following the discovery, the optical and infrared observations
of the optical counterpart to \gro~ revealed the presence of 
strong Balmer emission lines and infrared excess (Coe et al. 1994).
Based on these results, the system was classified as a massive binary 
system consisting of a neutron star as the compact object and a Be or 
a supergiant primary. After 260 days of this outburst, a second outburst was 
detected by BATSE (Finger et al. 1994). Assuming this 260 d as the orbital 
period of the pulsar, Finger et al. (1994) estimated the mass of the binary 
companion to be 3-8 $M_\odot$ indicating the system as a high mass X-ray binary. 
The ROSAT PSPC observation of the pulsar, in the declining phase of the discovery 
outburst by BATSE in 1993, clearly detected the 93.4 s pulsation with a 
double-peaked pulse profile in 0.1-2.4 keV light curve (Petre \& Gehrels 1994). 
A search in the archive of EXOSAT/Medium-Energy Experiment (ME) observation, 
centered on HD~88661, revealed the presence of the pulsed emission at the 
same period during the 1993 outburst (Macomb, Shrader, \& Schultz 1994). 
The X-ray spectrum (0.8--10 keV range) was found to be highly absorbed ($N_H
= 0.7\times10^{22} atoms cm^{-2}$) and described by a hard power-law with a 
photon index of $\sim$1.2. A combined analysis of data from the CGRO and ASCA
observations, though non-simultaneous, shortly after the peak of the discovery 
outburst, reported that the broadband spectrum of the pulsar can be well 
approximated by a power-law with an exponential cutoff and a 6.4 keV iron 
emission line (Shrader et al. 1999). The pulse profile was also found to be 
energy dependent, a double-peaked profile detected by ASCA that evolved into 
a single-peaked profile as detected by BATSE. Analyzing the BATSE and $Rossi 
X-ray Timing Explorer (RXTE)$/ASM flux histories, Shrader et al. (1999) suggested 
the orbital period of the system to be $\sim$135 days. However, Levine \& Corbet 
(2006) detected a 248.9 day periodicity in the $RXTE$ All-sky Monitor (ASM) 
X-ray light curve by analyzing data accumulated over nearly 10 years. This 
periodicity was found by visual identification of periodically occurring outbursts 
in the ASM light curve. An independent analysis of pulse period variations during
outbursts, using BATSE data, estimated the orbital period precisely to be 247.8 d 
(Coe et al. 2007) which is good agreement with the orbital period determined from 
the recurrence of the X-ray outbursts in ASM light curve.

Following the detection of an intense outburst from \gro~ with the Burst 
Alert Telescope (BAT) on Swift on 2007 November 17 (Krimm et al.\ 2007), the
accreting X-ray pulsar was observed with various X-ray observatories. The RXTE
observations detected the pulsar up to $\sim$70 keV along with the regular
$\sim$93.75 s pulsations (Wilms et al. 2007). Suzaku performed a TOO observation
of the pulsar on 2007 November 30. The results obtained from the analysis of the
Suzaku observation are presented in this paper.

\begin{figure}
\centering
\includegraphics[height=3.4in, width=3in, angle=-90]{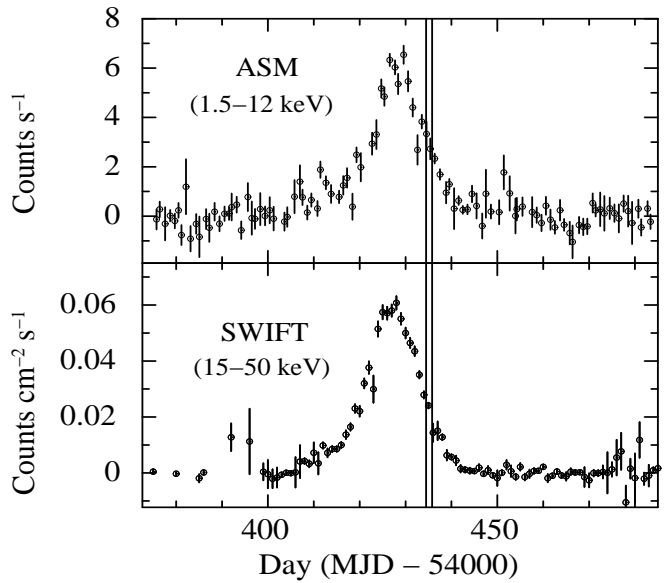}
\caption{The RXTE-ASM and Swift/BAT light curves of \gro\ in 1.5-12 keV 
and 15-50 keV energy bands, from 2007 September 29 (MJD 54372) to 2008 
January 20 (MJD 54485). The region between the vertical lines shows the 
duration of the Suzaku observation of the source.}
\label{asm}
\end{figure}

\section{Observation}

The transient pulsar \gro~ was observed with the $Suzaku$ during the declining 
phase of the 2007 November outburst. We used public data (ver-2.1.6.16) for 
the $Suzaku$ Target of Opportunity (TOO) observation of the pulsar in the present
work. The RXTE/ASM monitoring of the pulsar showed that the outburst lasted 
for $\sim$20 days. During this outburst, the peak luminosity was about $\sim$90
mCrab ($\sim$7 ASM counts s$^{-1}$). During this outburst, the pulsar was 
observed with the RXTE and $Suzaku$ observatories. The RXTE/ASM (1.5-12 keV) and 
Swift/BAT (15-50 keV) one-day averaged light curves of \gro~ between 2007 September 
29 and 2008 January 20 are shown in the top and bottom panels of Figure~\ref{asm}, 
respectively. The region between the vertical lines in the figure indicates the 
observation of the pulsar with the $Suzaku$. This TOO observation was carried out 
at ``XIS nominal'' pointing position for effective exposures of 42 ks. The XIS was 
operated with ``1/4 window'' option which gives a time resolution of 2 sec, covering 
a field of view of 17$'$.8$\times$4$'$.4.

$Suzaku$, the fifth Japanese X-ray astronomy satellite (Mitsuda et al. 2007),
was launched on 2005 July 10. It covers 0.2--600 keV energy range with the two
sets of instruments, X-ray Imaging Spectrometer (XIS; Koyama et al. 2007)
covering 0.2-12 keV energy range, and the Hard X-ray Detector (HXD; Takahashi 
et al. 2007) which covers 10--70 keV with PIN diodes and 30--600 keV with GSO
scintillators. Among the 4 sets of XISs, one is back illuminated (BI) whereas 
the other three are front illuminated (FI). The field of view of the XIS is 
18'$\times$18' in a full window mode with an effective area of 340 cm$^2$ (FI) 
and 390 cm$^2$ (BI) at 1.5 keV. The energy resolution was 130 eV (FWHM) at 6 
keV just after the launch. The HXD is a non-imaging instrument that is designed 
to detect high-energy X-rays. The HXD has 16 identical units made up of two 
types of detectors, silicon PIN diodes ($<$ 70 keV) and GSO crystal scintillator
($>$ 30 keV). The effective areas of PIN and GSO detectors are $\sim$ 145 
cm$^2$ at 15 keV and 315 cm$^2$ at 100 keV respectively. For a detailed
description of the XIS and HXD detectors, refer to Koyama et al. (2007) 
and Takahashi et al. (2007). As XIS-2 is no more operational, data from other 
3 XISs are used in the present analysis. The GSO data are used for timing
analysis here in this paper. As the pulsar is very faint at energies above 
50 keV, we did not use the GSO spectrum in the spectral fitting.

\section{Analysis and Results}

For XIS and HXD/PIN data reduction, we used the cleaned event data and obtained 
the XIS and PIN light curves and source spectra. The simulated background events,
as suggested by the instrument team, were used to estimate the HXD/PIN background 
for the \gro~ observation. The response file released in 2007 September was used 
for HXD/PIN spectrum. The accumulated events of the XIS data were discarded when 
the telemetry was saturated, the data rate was low, the satellite was in the South 
Atlantic Anomaly, and the source elevation above the Earth's limb was below 5$^\circ$ 
for night-Earth and below 20$^\circ$ for day-Earth. Applying these conditions, 
the source spectra were accumulated from the XIS cleaned event data by selecting 
a circular region of $4.3'$ around the image centroid. Because this extraction 
circle is larger than the optional window, the effective extraction region is 
the intersection of the window and this circle. The XIS background spectra were 
accumulated from the same observation by selecting rectangular regions away from 
the source. The response files and effective area files for XIS were generated by 
using the ''xissimarfgen'' and ''xisrmfgen'' tasks of FTOOLS (ver-6.3.1). The HXD/GSO
data was reprocessed to extract the source light curves for various energy ranges.
X-ray light curves of 2 s, 1 s, and 8 s time resolutions were extracted from the XIS, 
PIN, and GSO event data, respectively.

\begin{figure}
\centering
\includegraphics[height=2.75in, width=4.25in, angle=-90]{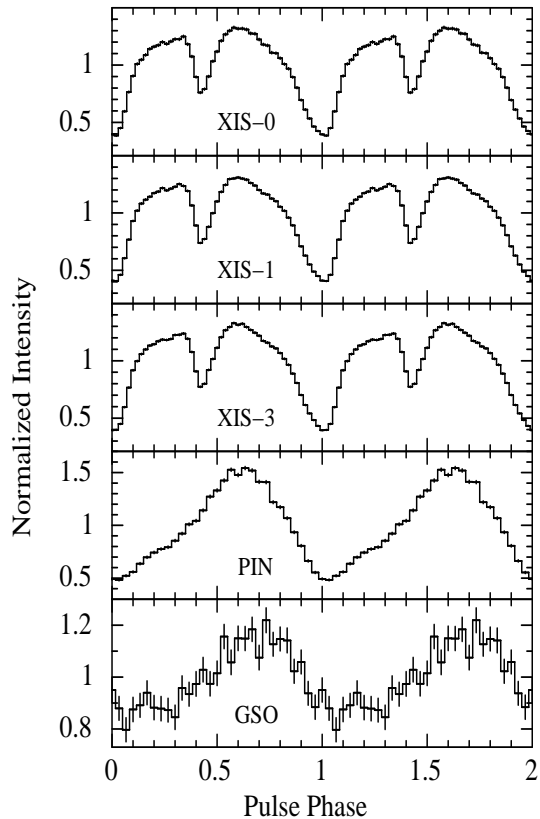}
\caption{The three XISs (in 0.2--12 keV energy band; top three panels), 
PIN (in 10--70 keV energy band; fourth panel) and GSO (in 50--200 keV 
energy band; bottom pannel) pulse profiles, obtained by using the estimated
pulse period of \gro~ for the $Suzaku$ TOO observation. Two pulses are shown 
for clarity. The errors in the figure are estimated for the 1~$\sigma$ 
confidence level.} 
\label{pp}
\end{figure}

\subsection{Timing Analysis}

\begin{figure*}
\vskip 14.2 cm
\includegraphics{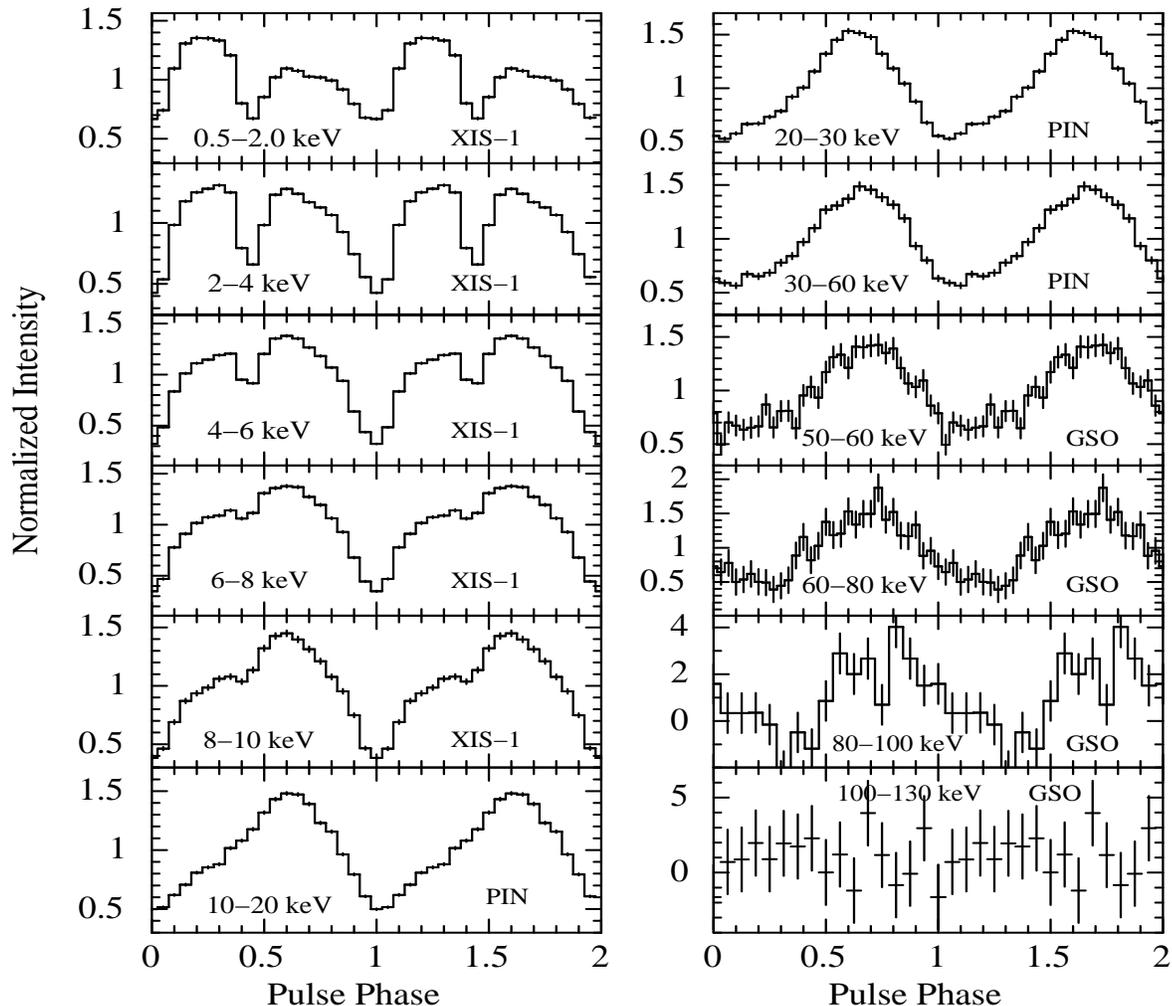}
\caption{The XIS, HXD/PIN and HXD/GSO pulse profiles of \gro~ at different
energy bands. We can see the presence/absence of the dip like structure in 
0.35-0.45 pulse phase range. The error bars represent 1~$\sigma$ uncertainties. 
Two pulses in each panel are shown for clarity.}
\label{erpp}
\end{figure*}

For the timing analysis, a barycentric correction was applied to the arrival 
times of the X-ray photons using the ``aebarycen'' task of FTOOLS. As described
above, light curves with a time resolutions of 2~s, 1~s and 8~s were extracted 
from XIS (0.2--12~keV), HXD/PIN (10--70~keV), and HXD/GSO (50--200 keV) event 
data. Pulse folding and a $\chi^2$ maximization method was applied to all the 
light curves yielding the pulse period of the pulsar to be 93.737(1) s. The pulse 
profiles obtained from the background subtracted XIS, HXD/PIN and HXD/GSO 
light curves of the $Suzaku$ observation of the pulsar are shown in Figure~\ref{pp}. 
The epoch was adjusted to obtain the minimum in the profile at phase zero. From 
the figure, it is observed that the shape of the pulse profiles obtained from the 
three XISs light curves are identical, whereas it is different to that obtained 
from the PIN and GSO data. The dip in the pulse profile in pulse phase range 
0.95-1.05, hereafter referred to as a primary dip, is narrow at soft X-rays 
(XIS energy band) and broad in the hard X-ray energy bands. Apart from the 
variable width of the dip, another dip like structure is present in the soft 
X-ray pulse profile in 0.35-0.45 pulse phase range which is absent in the hard 
X-ray profiles. To investigate the energy dependence of the pulse profile of 
\gro, we generated light curves in different energy bands from XIS, PIN and 
GSO event data. The background subtracted light curves are folded with the 
pulse period of the pulsar and the corresponding pulse profiles are shown in 
Figure~\ref{erpp}. To determine up to what energy pulsations 
are detected in GSO, we have searched for pulsations in narrow energy bands of 10 and
20 keV within the entire GSO energy range. The pulsations are clearly detected in the 
80-100 keV band, with a detection significance of 7.7$\sigma$. The background subtracted
light curves do not show any excess above 100 keV, and as expected, no pulsations are
detected beyond this  energy. The dip like structure (in pulse phase 
range 0.35-0.45) is found to be very prominent up to $\sim$4 keV. The width and depth 
of this structure decrease gradually with energy up to $\sim$10 keV, beyond which it
becomes indistinguishable in the pulse profile, making the hard X-ray pulse profiles
smooth and single peaked. Pulse phase resolved spectroscopy would help in understanding
the nature of the energy dependence of the dip like structure in the pulse profile 
of \gro.

\subsection{Spectral Analysis}

\subsubsection{Pulse phase averaged spectroscopy}

\begin{figure}
\centering
\includegraphics[height=2.85in, width=2.2in, angle=-90]{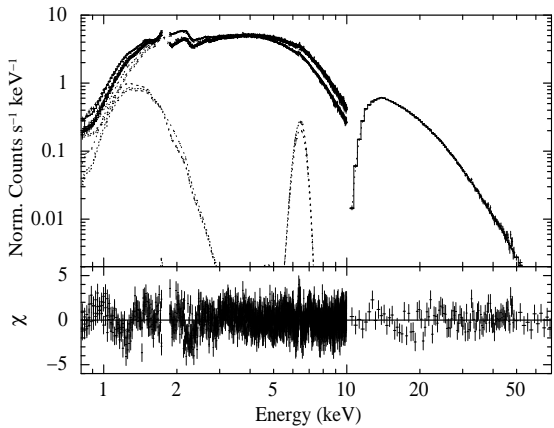}
\caption{Energy spectrum of \gro~ obtained with the XIS and PIN detectors 
of the $Suzaku$ TOO observation, along with the best-fit model comprising 
a blackbody component (BB), high energy cutoff power-law continuum model 
and a narrow iron line emission. The bottom panel shows the contributions 
of the residuals to the $\chi^2$ for each energy bin for the best-fit model.}
\label{spec-fg1}
\end{figure}

The broad-band spectrum of accretion powered X-ray pulsars are generally
described by a power-law, broken power-law or power-law with high energy
cutoff continuum models. In some cases, the pulsar spectrum has also been
described by the NPEX continuum model (Mihara 1995; Makishima et al. 1999;
Terada et al. 2006; Naik et al. 2008 and references therein). In case of a 
few X-ray pulsars, it has been reported that the absorption has two different 
components (Endo et al. 2000; Mukherjee \& Paul 2004). In this model, one 
absorption component absorbs the entire spectrum where as the other 
component absorbs the spectrum partially. This model is known as partial
absorption model. The choice of appropriate continuum model, therefore, is 
very much important to understand the properties of the pulsars. We tried 
to find a suitable continuum model to fit the broad-band spectrum of \gro.
In the process, we attempted to fit the pulsar spectra with different continuum
models with additional spectral components such as blackbody radiation,
iron emission line etc. The models giving statistically acceptable parameters
are described here in this section.

The source spectrum was extracted from the events selected in the energy 
ranges of 0.3-10.0 keV for the back illuminated CCD (XIS-1), 0.5-10 keV
for the front illuminated CCDs (XISs-0 and 3) and 10-70 keV for the HXD/PIN
detectors. After appropriate background subtraction, simultaneous spectral 
fitting was done using the XIS and PIN spectra with XSPEC V12. All the spectral
parameters other than the relative normalization, were tied together for all 
the detectors. Because an artificial structure is known to exist in the XIS
spectra around the Si edge, we ignored energy bins between 1.75--1.85~keV in 
the spectral analysis. Apart from the Si edge, large fit residuals due 
to calibration uncertainties are often observed near the edge structures of 
the XIS/XRT instrumental responses. Therefore, additional model components
for possible fluorescence lines at energies below $\sim$3.5 keV are not
considered during the spectral fitting.

\subsubsection{High energy cutoff power-law model}

We tried to fit the broad-band energy spectra (in 0.8-70 keV energy range) with 
a model consisting a power-law with exponential cutoff continuum along with the
interstellar absorption, and a Gaussian function for the iron fluorescence line 
at 6.4 keV. This model gave a reduced $\chi^2$ of $\geq$1.8 (for 1591 dof). 
The presence of positive residuals at the soft X-ray energy ranges allowed us to
add a blackbody component to the spectral model. The new model, when fitted with
to the 0.8-70 keV spectra improved the spectral fitting significantly yielding
a reduced $\chi^2$ of 1.47 (for 1589 dof). The relative instrument normalizations 
of the three XISs and PIN detectors were kept free and the values are found to be 
1.00:1.04:0.99:0.96 for XIS3:XIS0:XIS1:PIN with a clear agreement with that at 
the time of detector calibration. 

\begin{table*}
\centering
\caption{Best-fit parameters of the phase-averaged spectra for \gro~
during 2007 $Suzaku$ TOO observation with 1$\sigma$ errors. Model-1 : High energy 
cutoff power-law model with blackbody and Gaussian components, Model-2 : NPEX model 
with blackbody and Gaussian components, Model-3 : Partial covering high energy cutoff 
power-law model with Gaussian component}
\begin{tabular}{llll}
\hline
Parameter      		&\multicolumn{3}{|c|}{Value} 	 \\
				 &Model-1	&Model-2                  &Model-3\\
\hline
N$_{H1}$ (10$^{22}$ atoms cm$^{-2}$) &1.75$\pm$0.02	&1.49$\pm$0.01   &1.29$\pm$0.01\\
N$_{H2}$ (10$^{22}$ atoms cm$^{-2}$) &-----		&-----		 &5.0$\pm$0.1\\
Covering Fraction                    &-----		&-----		 &0.36$\pm$0.01\\
$kT_{BB}$ (keV)                   &0.17$\pm$0.01        &0.21$\pm$0.01   &-----\\
Iron line Energy (keV)            &6.46$\pm$0.01        &6.42$\pm$0.01   &6.39$\pm$0.01\\
Iron line width  (keV)            &0.27$\pm$0.01	&0.25$\pm$0.01   &0.07$\pm$0.02\\
Iron line eq. width (eV)          &61$\pm$7		&59$\pm$6        &21$\pm$4\\
Iron line flux$^a$                &8.9$\pm$0.8		&8.8$\pm$0.3     &3.2$\pm$0.4\\
Power-law index ($\alpha_1$)      &0.8$\pm$0.1		&0.20$\pm$0.01   &1.0$\pm$0.1\\
High energy cutoff (keV)	  &5.38$\pm$0.06	&6.8$\pm$0.1     &6.5$\pm$0.1\\
E-fold energy (keV)		  &17.3$\pm$0.1		&-----           &20.4$\pm$0.2\\
Blackbody flux$^a$                &5.76$\pm$0.35	&7.2$\pm$0.4     &-----\\
Abs. corrected BB flux$^b$	  &9.5$\pm$0.5		&5.7$\pm$0.4	 &-----\\
Reduced $\chi^2$                  &1.47 (1589)		&1.49 (1589)     &1.35 (1589)\\
\hline
\end{tabular}
\\
$^a$ : in 10$^{-12}$  ergs cm$^{-2}$ s$^{-1}$ unit, $^b$ : in 10$^{-11}$  ergs cm$^{-2}$ 
s$^{-1}$ unit, N$_{H1}$ = Equivalent hydrogen column density, N$_{H2}$ = Additional
hydrogen column density. Quoted source flux is not corrected for interstellar absorption.
\label{spec_par}
\end{table*}

We then tried to explore other continuum models to get a better fit to the broad 
band spectrum of \gro. In the process, we found that (i) the NPEX continuum model 
with blackbody and Gaussian components and (ii) partially covering high energy 
cutoff power-law model fit the pulsar spectrum with statistically acceptable 
parameters as the previous continuum model. These two additional models and 
results obtained from spectral fitting are described as follows.

\subsubsection{The NPEX continuum model} 

\begin{figure}
\centering
\includegraphics[height=2.85in, width=2.2in, angle=-90]{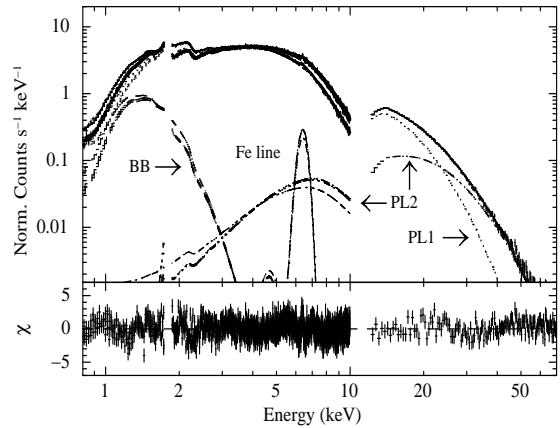}
\caption{Energy spectrum of \gro~ obtained with the XIS and PIN detectors 
of the $Suzaku$ TOO observation, along with the best-fit model comprising 
a blackbody component (BB), Negative Positive Exponential (NPEX) 
continuum model and a narrow iron line emission (Fe line). The negative 
and positive power laws are marked by PL1 and PL2 respectively. The bottom 
panel shows the contributions of the residuals to the $\chi^2$ for each energy 
bin for the best-fit NPEX continuum model.}
\label{spec-fg2}
\end{figure}

We found that, as in case of A0535+262 (Terada et al. 2006, Naik et al. 2008), 
the NPEX continuum model also describes the broad-band spectrum of \gro~ as well 
as the previous model. The NPEX continuum model which is an approximation of the 
unsaturated thermal Comptonization in hot plasma (Makishima et al. 1999). This 
continuum model reduces to a simple power-law with negative slope at low energies 
that is used to describe the spectra of accretion powered X-ray pulsars at low 
energies. The analytical forms of the NPEX model is

\begin{eqnarray}
\nonumber 
NPEX(E) = (N_1 E^{-\alpha_1} + N_2 E^{+\alpha_2})  ~ \exp \left( -\frac{E}{kT} \right)
\label{eq1}
\end{eqnarray}

where $E$ is the X-ray energy (in keV), $N_1$ and $\alpha_1$ are the
normalization and photon index of the negative power-law respectively,
$N_2$ and $\alpha_2$ are those of the positive power-law, and $kT$ is the
cutoff energy in units of keV. All five parameters of the NPEX continuum
component were kept free. As in the case of the $Suzaku$ observation of
the Be transient X-ray pulsar A0535+262 (Naik et al. 2008), we could not 
constrain the positive power-law index $\alpha_2$ at 2 and fixed it at
3 in the subsequent analysis. The relative instrument normalizations of 
the three XISs and PIN detectors were kept free and the values are found 
to be 1.00:1.03:0.99:1.00 for XIS3:XIS0:XIS1:PIN with a clear agreement 
with that at the time of detector calibration. The reduced $\chi^2$ obtained
from the simultaneous spectral fitting of the XIS and PIN data with the 
NPEX continuum model is found to be 1.49 (for 1589 dof). This is found to 
be comparable to the value obtained from the high energy cut-off power-law 
model.

\begin{figure}
\centering
\includegraphics[height=2.85in, width=2.2in, angle=-90]{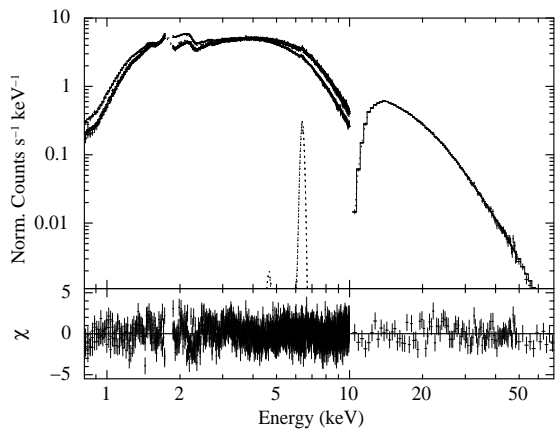}
\caption{Energy spectrum of \gro~ obtained with the XIS and PIN detectors 
of the $Suzaku$ TOO observation, along with the best-fit model comprising 
a partially absorbed high energy cutoff power-law  continuum model and a 
narrow iron line emission. The bottom panel shows the contributions of the 
residuals to the $\chi^2$ for each energy bin for the best-fit model.}
\label{spec-fg3}
\end{figure}

\subsubsection{Partial covering high energy cutoff power-law model}

The partial covering model is also known to be consisting of two power-law
continua with a common photon index but with different absorbing hydrogen
column densities. The analytical form of the partially covering high energy 
cutoff power-law model is

\begin{eqnarray}
\nonumber 
N(E) & = & {e^{-\sigma(E)N_{\mathrm H1}}}(S_{1}+S_{2}e^{-\sigma(E)N_{\mathrm H2}})
{E^{-\Gamma}}{I(E)}
\end{eqnarray}

where
\begin{eqnarray}
\nonumber  I(E) & = & 1  \hspace{0.83in} for ~E < E_\mathrm c \\
\nonumber       & = & e^{- \left({E-E_\mathrm c}\over{E_\mathrm f}\right)} \hspace{0.27in}  for ~E > E_\mathrm c
\end{eqnarray}

N(E) is the observed intensity, $\Gamma$ is the photon index, $N_{H1}$ 
and $N_{\mathrm H2}$ are the two equivalent hydrogen column densities, $\sigma$ 
is the photo-electric cross-section, $S_{1}$ and $S_{2}$ are the respective 
normalizations of the power law, $E_{\mathrm c}$ is the cut--off energy and 
$E_{\mathrm f}$ the e--folding energy. As in previous two cases, the relative 
instrument normalizations of the three XISs and PIN detectors were kept free. The 
values obtained are found to be 1.00:1.03:0.99:0.99 for XIS3:XIS0:XIS1:PIN with a 
clear agreement with the previous values. It is found that, unlike the previous 
two continuum models, the blackbody component for soft excess in the pulsar was 
not required to fit the 0.8--70 keV spectrum. The values of the equivalent column
densities N$_{H1}$ and N$_{H2}$ are found to be $\sim$1.3 $\times$ 10$^{22}$ atoms 
cm$^{-2}$ and 4.96 $\times$ 10$^{22}$ atoms cm$^{-2}$, respectively. The covering
fraction of the more absorbed power-law [= Norm$_2$ / (Norm$_1$ + Norm$_2$) = 
$S_2 / (S_1 + S_2)$] is found to be $\sim$0.36. The partial covering model showed 
improvement in the spectral fitting compared to the previous two continuum models 
with reduced $\chi^2$ of 1.35 (for 1589 dof).

The parameters of the three different continuum models obtained from the 
simultaneous spectral fitting to the XIS and PIN data of the $Suzaku$ observations 
of \gro~ are given in Table~\ref{spec_par}. The count rate spectra of the pulsar 
\gro~ are shown in Figure~\ref{spec-fg1} (for high energy cutoff power-law model), 
Figure~\ref{spec-fg2} (for NPEX model), and Figure~\ref{spec-fg3} (for partial 
covering model) along with the model components (top panels) and residuals to the 
best-fit model (bottom panels). 

The spectral fitting of the non-simultaneous OSSE and BATSE observations of 
the pulsar showed a marginal cyclotron resonance feature centered at $\sim$88 
keV (Shrader et al. 1999). The corresponding magnetic field of the pulsar is
estimated to be $\sim$10$^{13}$ G which is at the higher end of the magnetic 
field strengths of the neutron stars in the accreting X-ray binary pulsars, 
suggesting the detection at $\sim$88 keV could be the second harmonics. However, 
in our spectral fitting, no such absorption feature was present at $\sim$44 keV.
Therefore, we did not add any additional cyclotron absorption component to the 
spectral fitting at $\sim$44 keV.

\subsubsection{Pulse phase resolved spectroscopy}

The presence of energy dependent dips in the pulse profile of \gro~
prompted us to make a detailed study of the spectral properties
at different pulse phases of the transient pulsar. To investigate the 
changes in the spectral parameters at different pulse phases, we used 
data from the XIS (both BI and FIs) and HXD/PIN detectors. The XIS and 
PIN spectra were accumulated into 20 pulse phase bins by applying phase 
filtering in the FTOOLS task XSELECT. The XIS and PIN background spectra 
and response matrices used for the phase averaged spectroscopy, were also 
used for the phase resolved spectroscopy. Simultaneous spectral fitting 
was done in the 1.0-70.0 keV energy band. The phase resolved spectra were 
fitted with all the three continuum models used to describe the phase averaged 
spectrum, separately. We fixed some of the parameters such as the values of
blackbody temperature (in case of Model-1 and Model-2, as given in Table-1), 
absorption column density, iron line energy and line at the phase averaged 
values. The relative instrument normalizations were fixed at the values 
obtained from the phase averaged spectroscopy. It was found that Model-1 and
Model-2, when used to fit the phase resolved spectra, did not provide good 
fit for some of the pulse phases. It was also found that the blackbody flux
(i.e. the soft excess) was significantly high during the dip like feature 
in the pulse profile. It is difficult to explain a significant increase in 
the blackbody flux during a narrow pulse phase (during the dip like feature 
in the pulse profile). This is our main basis to reject the first two models 
and use the partial covering model (Model-3 -- as given in Table-1) for pulse 
phase resolved spectroscopy. All the 20 pulse phase resolved spectra were 
fitted well with this model. 

\begin{figure*}
\centering
\includegraphics[height=5.5in, width=4.5in, angle=-90]{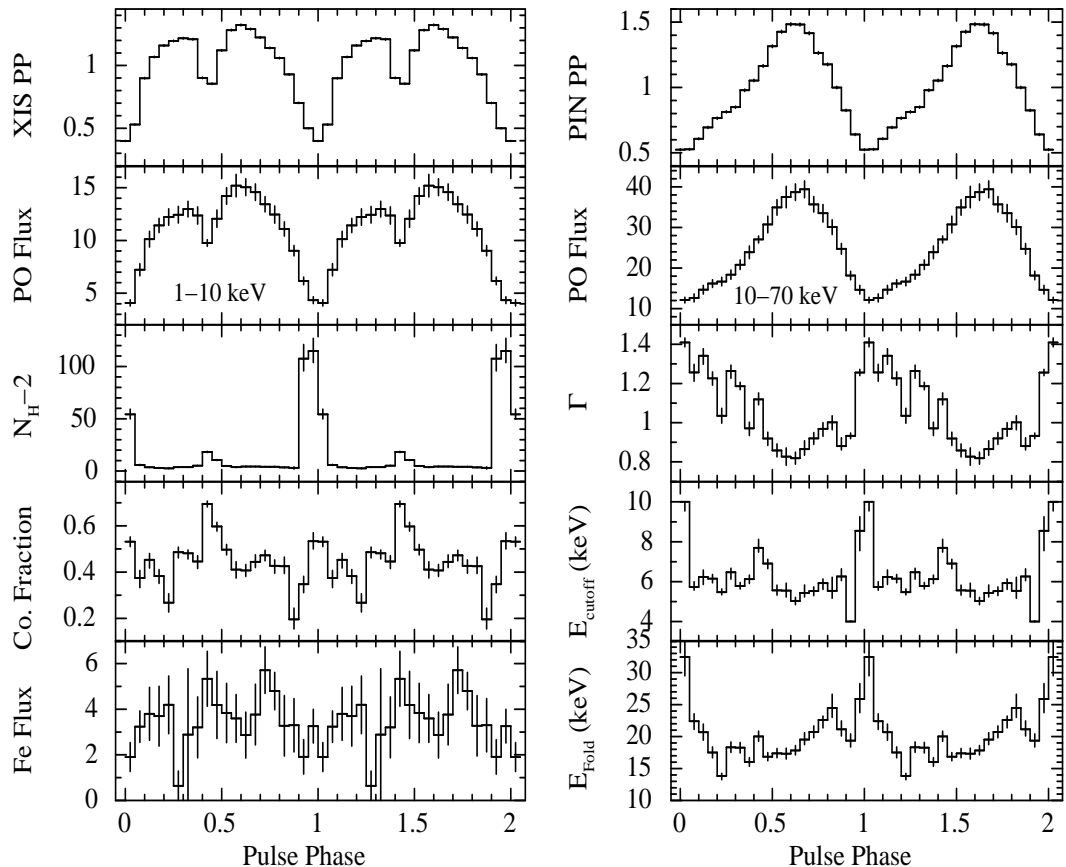}
\caption{Spectral parameters obtained from the pulse phase resolved
spectroscopy of $Suzaku$ observation of \gro. The errors shown
in the figure are estimated for 1$\sigma$ confidence level. In the
figure, the iron line flux (Fe Flux), power-law flux (PO Flux), and
N$_{H2}$ are plotted in the units of 10$^{-12}$ ergs cm$^{-2}$ 
s$^{-1}$, 10$^{-10}$ ergs cm$^{-2}$ s$^{-1}$, and 10$^{22}$ atoms cm$^{-2}$,
respectively. The power-law flux is not corrected for low-energy absorption. 
The XIS and PIN pulse profiles are shown in the left top and right top panels 
respectively.}
\label{phrs}
\end{figure*}

The parameters obtained from the simultaneous spectral fitting to the XIS 
and PIN phase resolved spectra with Model-3 are shown in Figure~\ref{phrs}. 
The top panels of the figure shows the pulse profiles of the pulsar obtained 
from XIS (left panel) and PIN data (right panel). The second panels show 
estimated source flux in 1-10 keV (left panel) and 10-70 keV (right panel). 
It can be seen that the shape of the pulse profiles and the estimated source 
flux over the pulse phases follow exactly the same pattern. It can be seen
that the value of N$_{H2}$ is significantly high in 0.9--1.05 pulse phase 
range (the primary dip in the pulse profile). Very large value of N$_{H2}$ 
in the above pulse phase range most probably due to the presence of accretion
column or accretion stream in the pulsar. The high values of N$_{H2}$ with 
high values of covering fraction at the pulse phase of dip like feature in the 
pulse profile (0.4--0.5 phase range) suggest that the dip like feature is caused 
by the absorption due to additional matter at that pulse phase range. The 
indifferent values of the iron emission line flux suggests that the matter 
emitting the iron fluorescence line is probably distributed symmetrically 
around the pulsar. The values of the power-law photon index, cut-off energy, 
and e-folding energy are found to be maximum in 0.9-1.1 pulse phase range i.e. 
at the main dip in the pulse profile. There is no remarkable change in the values 
of above three parameters during the short dip in the pulse profile 
(in 0.4--0.5 pulse phase range).

\section{Discussion}

\subsection{Pulse Profile}
The temporal and spectral properties of the Be transient X-ray pulsar
\gro~ have been reported only in a couple of occasions since its discovery.
We detected X-ray pulsations in GRO~J1008-57 as high as the 80-100 keV
energy band, which has not been reported earlier. The pulse profile of \gro~ 
is found to be strongly energy dependent i.e. a double-peaked profile in the 
soft X-ray energy band ($<$ 10 keV) and a single-peaked smooth profile in 
hard X-rays. The double-peaked
profile at soft X-rays, as shown here in this paper, is found to be different
from that of the single-peaked profile (in 1--4 keV energy range) obtained 
from the EXOSAT observation of \gro~ in 1985 (Macomb et al. 1994). Similar
changes in the shape of pulse profiles are also seen in the Be transient 
X-ray pulsar A0535+262 (Naik et al. 2008), a single peaked profile in 
the quiescence (Mukherjee \& Paul 2005) and a double-peaked profile during 
outbursts (at high luminosity). The double-peaked pulse profiles seen in \gro, 
agrees with the luminosity dependence of the pulse profiles as seen in other
pulsars. The presence of dip like structures in the pulse profiles of these 
X-ray pulsars is described as due to the obscuration of matter to the radiation. 
As we showed here, the dip like feature in the pulse profile of \gro~ is 
probably due to the additional absorption (other than the Galactic column 
density) at the pulse phase. 

\subsection{Spectroscopy}
The broad-band X-ray spectrum of \gro~ has been described here for the 
first time in detail. Shrader et al. (1999) tried to explain the pulsar 
spectrum obtained from the OSSE data by a thermal bremsstrahlung model 
with a characteristic temperature $kT$ = 19 keV. However, when the 
spectral fitting was extended towards the low energy (ASCA energy range), 
the fitting was inconsistent. The statistics was improved marginally because 
of the addition of an absorption component at $\sim$88 keV for the possible 
cyclotron features. Shrader et al. (1999), when fitted the ASCA, BATSE and 
OSSE data simultaneously, found that the broad-band spectrum can be described 
by an absorbed power-law with high energy cutoff and a Gaussian component for 
the Iron emission line. However, the ASCA, BATSE, and OSSE observations of the
pulsar were not simultaneous. The midpoints of ASCA and OSSE observations used 
were separated by 4 days during which time the source intensity was halved.
As the spectrum of Be transient pulsars can differ at different luminosity
states, the broad-band spectral analysis of the Suzaku observations of the 
pulsar can give better information on the properties. 

Selection of an appropriate continuum model is important to investigate
the presence of several features in the pulsar spectrum, such as the soft 
excess represented by a blackbody component, emission lines, cyclotron 
absorption features etc. Most of the transient Be X-ray binary pulsars 
undergo periodic outbursts due to the enhanced mass accretion when the 
neutron star passes through the dense regions of the circumstellar disk
or periastron passage when the outer edge of the disk is stripped off
resulting in sudden accretion of matter onto the neutron star. 
During the passage, the value of the absorption column density increases 
compared to the value of the Galactic column density in the source direction. 
The spectral fitting to the data obtained from ASCA, BATSE, OSSE observations 
of the transient pulsar \gro~ during the 1993 August outburst, yielded the 
value of $N_H$ in the range of 0.8--1.73 $\times$ 10$^{22}$ atoms cm$^{-2}$ 
(Table~1; Shrader et al. 1999). In case of data obtained from the Suzaku 
observation of the pulsar, it is found that the 0.8-70.0 keV broad-band spectra 
can be well described by three different continuum models with similar statistical 
parameters. The high energy cutoff power-law model and NPEX continuum model 
yielded higher value of $N_H$ than that of the Galactic value in the direction 
of the pulsar. It is interesting to note that, inspite of a high value of column 
density, a blackbody component of temperature $kT$ $\sim$0.2 keV was also required 
for these two continuum models to describe the broad-band spectrum of the pulsar. 
In these two models, it is estimated that the absorption corrected flux of the 
soft X-ray excess (blackbody) in \gro~ is about 2\% of the unabsorbed source flux 
in 0.8-70 keV energy range. 

The third model i.e. the partial covering model, however, fits the pulsar spectrum 
comparatively better than the previous two continuum models. Based on our results 
from the phase resolved spectroscopy, the earlier two continuum models were not 
preferred to describe the properties of the pulsar. In the partial covering model,
N$_{H1}$ is considered as the Galactic hydrogen column density, and N$_{H2}$ is 
interpreted as the column density of the material that is local to the neutron star. 
The value of N$_{H2}$ is maximum during the primary dip that is interpreted as due 
to the accretion column. The high value of N$_{H2}$ and the covering fraction at 
0.4--0.5 pulse phase range explain the dip like feature in the pulse profile.
The broad-band spectroscopy of \gro~ also shows the presence of a narrow iron 
K$_\alpha$ emission line at 6.4 keV. The iron emission line is generally interpreted 
as due to the fluorescent line from the cold matter in the surrounding region of 
the neutron star.

\section*{Acknowledgments}
We wish to thank the referee for his/her suggestions on the paper. The 
research work at Physical Research Laboratory is funded by the Department of
Space, Government of India. SN thanks Zulfikar Ali for useful discussion in
HXD/GSO data processing. CK would like to acknowledge the hospitality provided 
by the Physical Research Laboratory during his visit to carry out the present 
work. The authors would like to thank all the members of the Suzaku for their 
contributions in the instrument preparation, spacecraft operation, software 
development, and in-orbit instrumental calibration. This research has made use 
of data obtained through HEASARC Online Service, provided by the NASA/GSFC, in 
support of NASA High Energy Astrophysics Programs.


\begin{thebibliography}{}
\bibitem[]{}Coe, M. J., Roche, P., Everall, C., et al. 1994, MNRAS, 270, L57
\bibitem[]{}Coe, M. J. 2000, ASPC, 214, 656
\bibitem[]{}Coe, M. J., Bird, A. J., Hill, A. B., McBride, V. A., Schurch, M.,
            Galache, J., Wilson, C. A., Finger, M., Buckley, D. A.,
            Romero-Colmenero, E. 2007, MNRAS, 378, 1427
\bibitem[]{}Corbet, R. H. D. 1986, MNRAS, 220, 1047
\bibitem[]{}Finger, M. H., Wilson, R. B., Scott, M., Stollberg, M., 
	    Prince, T. A. 1994, IAU Circ. 5977
\bibitem[]{}Finger, M. H., Wilson, R. B., Chakrabarty, D. 1996, A\&AS, 120, 209
\bibitem[]{}Hickox, R. C., Narayan, R., \&  Kallman, T. R. 2004, ApJ, 614, 881
\bibitem[]{}Koyama, K., et al. 2007, PASJ, 59, 23
\bibitem[]{}Krimm, H. A., et al. 2007, Astron. Telegram, 1298
\bibitem[]{}Levine, A. M. \& Corbet, R. 2006, Astron. Telegram, 940
\bibitem[]{}Macomb, D. J., Shrader, C. R., \& Schultz, A. B. 1994, ApJ, 437, 845
\bibitem[]{}Makishima, K. 1986, in Physics of Accretion onto Compact Objects, ed. 
	    K. O. Mason, M. G. Watson, \& N. E. White (Berlin: Springer), 249
\bibitem[]{} Makishima, K., Mihara, T., Nagase, F., \& Tanaka, Y. 1999,
ApJ, 525, 978
\bibitem[]{}Massi, M., Ribo, M., Paredes, J. M., et al. 2004, A\&AL, 414, 1 
\bibitem[]{}Mihara, T. 1995, Ph.D. thesis, Univ. Tokyo
\bibitem[]{}Mitsuda, K., et al. 2007, PASJ, 59, 1
\bibitem[]{}Mukherjee, U. \& Paul, B. 2005, A\&A, 431, 667
\bibitem[]{}Mukherjee, U. \& Paul, B. 2004, A\&A, 427, 567
\bibitem[]{}Naik, S., et al. 2008, ApJ, 672, 516
\bibitem[]{}Naik, S., Callanan, P. J., Paul, B., \& Dotani, T. 2006, ApJ, 647
\bibitem[]{}Naik, S. \& Paul, B. 2004a, ApJ, 600, 351
\bibitem[]{}Naik, S. \& Paul, B. 2004b, A\&A, 418, 655
\bibitem[]{}Negueruela, I., Reig, P., Coe, M. J., \& Fabregat, J. 1998, 
	    A\&A, 336, 251
\bibitem[]{}Okazaki, A. T., Bate, M. R., Ogilvie, G. I., Pringle, J. E. 2002,
	    MNRAS, 337, 967
\bibitem[]{}Okazaki, A. T., Negueruela, I. 2001, A\&A, 377, 161
\bibitem[]{}Paul, B., Nagase, F., Endo, T., Dotani, T., Yokogawa, J.,
            Nishiuchi, M. 2002, ApJ, 579, 411
\bibitem[]{}Petre, R. \& Gehrels, N. 1994, A\&A, 282, L33
\bibitem[]{}Reig, P., Negueruela, I., Buckley, D. A. H., et al. 2001, 
	    A\&A, 367, 266 
\bibitem[]{}Shrader, C. R., Sutaria, F. K., Singh, K. P., and Macomb, D. J. 1999, 
	    ApJ, 512, 920
\bibitem[]{}Stollberg, M. T., Finger, M. H., Wilson, R. B., Harmon, B. A.,
	    Rubin, B. C., Zhang, N. S., Fishman, G. J. 1993, IAU Circ. 5836
\bibitem[]{}Takahashi, T., et al. 2007, PASJ, 59, 35
\bibitem[]{}Terada, Y., et al. 2006, ApJ, 648, L139
\bibitem[]{}Wilms, J., et al. 2007, Astron. Telegram, 1304
\bibitem[]{}
\end{thebibliography}
\end{document}